\documentclass[conference]{IEEEtran}
\usepackage[T1]{fontenc}
\usepackage{amsmath,amssymb,booktabs,array,graphicx,cite,balance}
\usepackage[hidelinks]{hyperref}
\title{An Evidence-Guided Framework for Tissue-Barrier Impedance Review: Deletion and Resistance Sensitivity}
\author{\IEEEauthorblockN{Leo Yang\IEEEauthorrefmark{1}, Yingtong Zhou\IEEEauthorrefmark{2}\IEEEauthorrefmark{3}}
\IEEEauthorblockA{\IEEEauthorrefmark{1}UWC South East Asia, 1207 Dover Rd, Level 2, Singapore 139654}
\IEEEauthorblockA{\IEEEauthorrefmark{2}IMT Atlantique, Technop\^ole Brest-Iroise, CS 83818, 29238 Brest Cedex 3, France}
\IEEEauthorblockA{\IEEEauthorrefmark{3}Corresponding author: yingtong.zhou@imt-atlantique.net}}
\begin{document}
\maketitle
\begin{abstract}
Reviewing tissue-barrier impedance requires connecting spectral anomalies with resistance estimates and experimental context. We present an evidence-guided framework that coordinates numerical checks and traceable actions for human review. In-house simulations assess the effect of removing flagged frequencies, public datasets test model adaptation, and local measurements demonstrate the review workflow. Removing flagged frequencies reduced resistance error under quadratic objectives, whereas robust fitting showed little average benefit from deletion and a mean error increase for endpoint-pair perturbations. High errors clustered around a few fitted circuit parameter sets, highlighting the value of examining parameter identifiability. Under the tested noise and perturbation conditions, these findings support robust full-spectrum estimation, flag-guided inspection, and matched sensitivity checks.
\end{abstract}

\section{Introduction}
Electrical impedance spectroscopy (EIS) monitors epithelial and endothelial barriers~\cite{Srinivasan2015,Huang2023,Liang2023}. Interpreting electrode, cell-layer, and acquisition effects requires metadata~\cite{Sharma2025} and separate checks of local deviations, circuit adequacy, and resistance reliability.

Linear Kramers--Kronig tests assess spectral consistency~\cite{Schonleber2014}; automated methods remove outliers~\cite{Zhao2023}; AutoEIS infers circuits~\cite{Zhang2023}; and AIA-EIS combines detection, circuit selection, and fitting~\cite{Zhao2026}. Fitting criteria affect parameter recovery~\cite{Jaberi2026}. Deletion influence measures sensitivity~\cite{Cook1977}, while robust M-estimation limits large-residual contributions~\cite{Huber1964}. Flagged-point removal therefore warrants evaluation when the estimator already downweights large residuals.

\begin{figure*}[t]
\centering
\includegraphics[width=.98\textwidth]{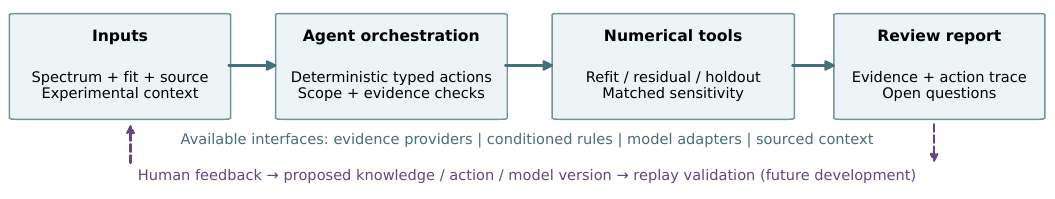}
\caption{Implemented review architecture. Deterministic policies connect evidence to numerical actions and traceable reports. Interfaces accommodate experimental context, knowledge rules, and numerical models. The dashed path denotes future user-curated extensions and replay validation.}
\label{fig:framework}
\end{figure*}

This study quantifies how a fixed flag set changes resistance error across three fitting objectives. A shared quadratic/pseudo-Huber scale enables a controlled comparison of penalty effects; common initialization controls the paired comparisons. Location strata reveal where deletion helps or harms. A deterministic review framework (Fig.~\ref{fig:framework}) implements the resulting inspection and sensitivity actions, connecting numerical evidence with experimental context.

\section{Methods}
\subsection{Datasets}
Table~\ref{tab:data} summarizes the datasets. Public epithelial-tissue simulations~\cite{Schindler2021} pair control/Nystatin with clean/scattered variants on 42 frequencies. We evaluate 30 simulated tissue units (four spectra each) from an existing 30/30-unit development/evaluation split. Relative LS fits a terminal resistance and two parallel-RC branches in series. Public organ-on-chip measurements from primary lung-cell cultures use 28 frequencies and $R_s+[R_{ct}^{-1}+K(j\omega)^\alpha]^{-1}$~\cite{Marrero2025}.

The local cohort comprises seven ordered acquisition records from MesoViva SAS, France: three labelled ips-endo (31 spectra) and four labelled EC (251 spectra), spanning monitoring and drug-administration contexts. An MV-MIA-A1 applied 0.1~V at 65 frequencies (5~Hz--100~kHz). Per record of $N_{\mathrm{rec}}$ spectra, three fitted parameter vectors at rounded indices $(N_{\mathrm{rec}}-1)\{0.25,0.50,0.75\}$ define 21 anchor circuits for simulation ($R_c=48.0$--3354.1~$\Omega$). These records and perturbations were examined during development; all 282 spectra enter retrospective replay.

\begin{table}[t]
\caption{Data Sources and Evaluation Roles}
\label{tab:data}
\centering\footnotesize\setlength{\tabcolsep}{3pt}
\begin{tabular}{@{}>{\raggedright\arraybackslash}p{.30\columnwidth}>{\raggedright\arraybackslash}p{.28\columnwidth}>{\raggedright\arraybackslash}p{.33\columnwidth}@{}}
\toprule
Source & Spectra & Evaluation role\\
\midrule
In-house simulation & 1,134 injected; 126 controls & Known-truth parameter error\\
Public simulation~\cite{SchindlerData2021} & 240 total; 120 evaluated & DC-resistance recovery\\
Public measurements~\cite{Marrero2025} & 97 & Source-model fit and flag transfer\\
Local measurements & 282 (7 records) & Retrospective policy replay\\
\bottomrule
\end{tabular}
\end{table}

\subsection{Agent Orchestration and Review Actions}
The \emph{agent} maps spectra, fitted models, provenance, and context to predefined review actions. Deterministic selectors invoke numerical adapters or request context; optional language-model outputs supply advisory text. Reports retain evidence identifiers, input hashes, fitting settings, action histories, and open findings.

The policy (Table~\ref{tab:actions}) retains full-spectrum robust estimates and carries forward existing parameter-review requests as input annotations. Internal-flag sensitivity uses $100|\widehat R_{c,\mathrm{del}}/\widehat R_{c,\mathrm{full}}-1|$ with a 10\% engineering escalation threshold. Policy settings were fixed for retrospective replay; the 10\% global-fit threshold was selected after cohort inspection.

\begin{table}[t]
\caption{Evidence-to-Action Policy for Local Review}
\label{tab:actions}
\centering\footnotesize\setlength{\tabcolsep}{3pt}
\begin{tabular}{@{}>{\raggedright\arraybackslash}p{.21\columnwidth}>{\raggedright\arraybackslash}p{.43\columnwidth}>{\raggedright\arraybackslash}p{.27\columnwidth}@{}}
\toprule
Evidence & Action & Recorded output\\
\midrule
Endpoint flag & Retain point; request inspection & Annotation\\
Internal flag & Matched robust sensitivity; retain endpoints & Estimates, relative change\\
$E_m>10\%$ or existing parameter request & Preserve model/parameter review request & Evidence, further checks\\
Missing context & Request source-linked background & Unresolved scope\\
\bottomrule
\end{tabular}
\end{table}

\subsection{Circuit, Fit Quality, and Paired Objectives}
The local constant-phase-element (CPE)/bulk/parallel-RC model is~\cite{Liang2023}
\begin{equation}
Z(\omega;\theta)=[a(j\omega)^n]^{-1}+R_b+
\frac{R_c}{1+j\omega R_cC_c},\qquad \omega=2\pi f.
\label{eq:circuit}
\end{equation}
Here $\theta=(a,n,R_b,R_c,C_c)$; $a$ has units S\,s$^n$, $n$ is dimensionless, and $C_c$ is in F. We report bulk resistance $R_b$ and parallel-branch resistance $R_c$ in $\Omega$, before blank subtraction or area normalization. For observed complex impedances $Z_k$, global fit quality uses relative RMS,
\begin{equation}
E_m=\left[\frac{1}{N}\sum_{k=1}^{N}
\frac{|Z(\omega_k;\widehat\theta_m)-Z_k|^2}{|Z_k|^2}\right]^{1/2}.
\label{eq:residual}
\end{equation}
where $N$ is the frequency count and $m$ the fitting method. All observations have positive magnitude.

Let $D$ contain the union of the frozen local flags and $K$ contain either all frequencies or their complement to $D$. Define $r_k=Z(\omega_k;\theta)-Z_k$, $d_k=r_k/|Z_k|$, and $u_k=r_k/s_k$. The objectives for relative least squares (LS), fixed-scale quadratic loss, and radial pseudo-Huber loss are
\begin{equation}
\begin{split}
L_{\mathrm{LS}}&=\sum_{k\in K}|d_k|^2,\qquad
L_{\mathrm{Q}}=\sum_{k\in K}|u_k|^2,\\
L_{\mathrm{R}}&=\sum_{k\in K}2c^2\left(\sqrt{1+|u_k|^2/c^2}-1\right),
\quad c=2.5.
\end{split}
\label{eq:objectives}
\end{equation}
A full-spectrum LS pilot supplies predictions $\mu_k$ and a frequency-dependent working scale $s_k=\max(|\mu_k|,10^{-9})\sigma_k$. The relative noise estimate $\sigma_k$ uses robust aggregation of locally detrended pilot residuals across frequencies and within frequency bands. The same $s_k$ is shared by $L_{\mathrm{Q}}$ and $L_{\mathrm{R}}$ and held fixed for full/deleted fits for the paired penalty comparison. Relative LS also differs in weighting.

We generate six initial vectors from each full spectrum and reuse them, the parameter bounds, and the solver settings for every comparison arm. SciPy~\cite{Virtanen2020} optimizes log parameters; we retain the lowest-cost converged fit for each arm.

\begin{table*}[t]
\caption{Full and Flag-Deleted Resistance Error in 1,134 Injected Spectra}
\label{tab:deletion}
\centering\footnotesize\setlength{\tabcolsep}{2.8pt}
\begin{tabular}{@{}lccccc@{}}
\toprule
Objective & Mean $e$ (\%), full$\to$del. & Median $e$ (\%), full$\to$del. & Mean $\Delta e$ [95\% interval], pp & No. $e>10\%$ & Improved / worsened / within\\
\midrule
Relative LS & $6.20\to4.74$ & $4.02\to2.44$ & $-1.47\ [-2.04,-0.84]$ & $218\to150$ & 543 / 192 / 399\\
Fixed-scale quadratic & $8.69\to4.70$ & $5.64\to2.34$ & $-3.99\ [-5.37,-2.75]$ & $337\to153$ & 647 / 144 / 343\\
Pseudo-Huber & $4.63\to4.64$ & $2.54\to2.57$ & $+0.01\ [-0.41,0.61]$ & $150\to150$ & 281 / 219 / 634\\
\bottomrule
\end{tabular}
\par\smallskip\raggedright\scriptsize
Counts: improved, $\Delta e<-1$~pp; worsened, $\Delta e>1$~pp; within, $|\Delta e|\leq1$~pp. Intervals condition on seven records, resampled within groups. All selected fits converged. There were 148 empty masks; for the 986 nonempty masks, mean changes were $-1.69$, $-4.59$, and $+0.012$ pp for LS, quadratic, and robust fitting.
\end{table*}

\subsection{Local Markers and Calibration}
The local-trend marker (M1) predicts each impedance using a robust quadratic trend through the nine nearest other log frequencies, with one-sided neighborhoods at endpoints. Its score standardizes the complex prediction residual using robust real/imaginary scales. The model-residual marker (M5) scores residual magnitudes from a robust circuit fit, normalized by LS pilot magnitudes and frequency-dependent residual scales. M5 uses a separate three-start pilot/robust fit.

Calibration used two fixed synthetic circuit settings on 65 log-spaced frequencies (5~Hz--100~kHz). Independent Gaussian component noise had SD 2\% of latent magnitude or 1\%--5\% increasing linearly with log frequency; 15 realizations per setting/profile yield 60 spectra. Per marker, the 58th smallest of 60 spectrum-maximum scores sets the cutoff: approximately 5.031 (M1) and 4.498 (M5). Full-precision cutoffs are frozen. Their strict-exceedance union defines one full-spectrum flag set, held fixed across all comparison arms. Implementation settings and source provenance are recorded in the accompanying code documentation.

\subsection{Evaluation Protocol}
The perturbation protocol was frozen before generating 1,134 spectra and 126 controls on these same anchors. Independent Gaussian real/imaginary component noise has SD 1\%, 2\%, or 5\% of latent impedance magnitude. Two noisy observations are multiplied by 1.10, 1.25, or 1.50, preserving their phase. Zero-based location pairs are $(0,64)$, $(21,43)$, and $(0,32)$. Two noise realizations per anchor yield $21\times2\times3\times3\times3=1134$ injected spectra and $21\times2\times3=126$ noise-only controls. Matching conditions share unit noise (generation seed 2026101807). Generation and fitting use~\eqref{eq:circuit}. The fixed-scale quadratic arm was specified and run as a subsequent extension on these saved observations.

Scoring uses known simulation truth: $e=100|\widehat R_c/R_c^\star-1|$ and $\Delta e=e_{\mathrm{deleted}}-e_{\mathrm{full}}$ in percentage points (pp). We report means, medians, quantiles, and changes beyond $\pm1$~pp, with the remaining cases grouped separately. The 10\% error threshold is an engineering endpoint; the 1-pp reporting interval was selected post hoc. An oracle arm removes the two injected frequencies under LS and robust fitting. Conditional 95\% percentile intervals use 5,000 paired whole-record resamples within the two groups of three and four records, preserving each record's anchors and perturbations.

\section{Results}
\subsection{Deletion Benefit Depends on the Objective}
Table~\ref{tab:deletion} summarizes the injected-cohort results. Deletion reduced mean error by 1.47~pp under relative LS and 3.99~pp under fixed-scale quadratic fitting. Robust fitting showed a mean change of $+0.01$~pp. With the same full-data scale, quadratic full-spectrum error exceeded robust full-spectrum error by 4.06~pp (conditional interval $[2.87,5.41]$). The fixed-scale comparison supports a contribution of the pseudo-Huber penalty to robustness against spikes, consistent with the importance of objective selection~\cite{Jaberi2026}.

For the practical choice between LS after flag deletion and robust full-spectrum fitting, mean error differed by $+0.10$~pp (LS deleted minus robust full; conditional 95\% interval $[-0.36,0.61]$). The small observed difference supports robust full-spectrum fitting as a practical option; formal equivalence has not been established and requires a prespecified margin and design.

\begin{figure*}[t]
\centering
\includegraphics[width=.94\textwidth]{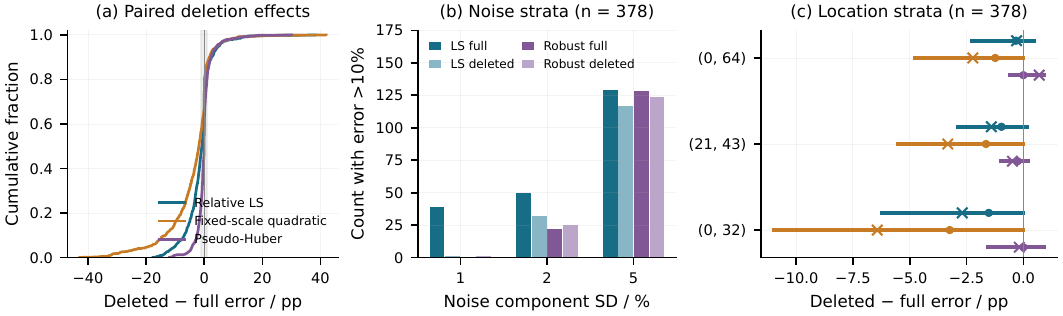}
\caption{Paired perturbation results. (a) Error-change distributions for all 1,134 injected spectra; shading marks $\pm1$ pp. (b) Counts above 10\% error, 378 spectra per noise level. (c) Injection-location strata, 378 cases each: bars show quartiles, circles medians, and crosses means. Indices 0 and 64 denote 5 Hz and 100 kHz.}
\label{fig:sensitivity}
\end{figure*}

Robust deletion moved 28 cases out of the $e>10\%$ category and 28 into it, leaving its total at 150. Its 5th--95th change percentiles were $[-4.14,4.38]$~pp (Fig.~\ref{fig:sensitivity}a), showing offsetting effects. Oracle deletion yielded 138 LS and 150 robust cases above 10\%.

\begin{figure*}[t]
\centering
\includegraphics[width=.94\textwidth]{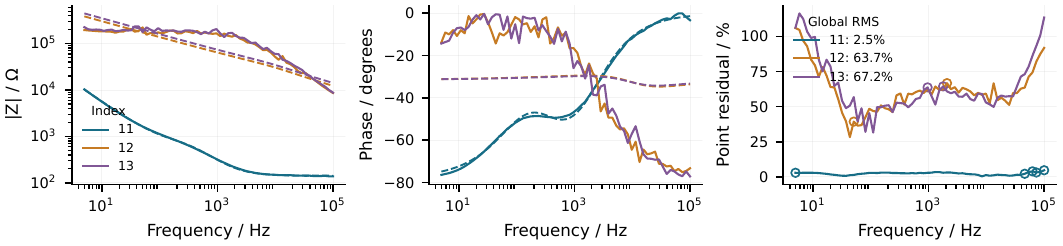}
\caption{Consecutive spectra from exp1\_0310 (zero-based indices 11--13): observations (solid), archived fits (dashed), and relative residuals with union flags (circles). Legends report global $E_m$. Index 11 illustrates policy replay; the final two spectra prompt acquisition-context review.}
\label{fig:local}
\end{figure*}

\subsection{Noise, Location, and Anchor Dependence}
At noise SDs of 1\%, 2\%, and 5\%, robust counts above 10\% error changed $0\to1$, $22\to25$, and $128\to124$ among 378 spectra per level; Fig.~\ref{fig:sensitivity}b compares LS. For both LS and robust fitting, noise-only counts were 0, 3, and 12 among 42 controls per level.

Robust mean $\Delta e$ was $+0.70$~pp for endpoint-pair injections, $-0.48$~pp for internal-pair injections, and $-0.18$~pp for endpoint/middle injections (Fig.~\ref{fig:sensitivity}c). These strata describe injection locations; deletion uses the complete detected mask. Oracle removal gave $+0.90$ and $-0.54$~pp for endpoint and internal pairs. Flags recovered 1,846/2,268 injected points (81.4\% recall), with 1,846/2,664 correct markers (69.3\% precision). All 126 controls had empty masks. At 5\% noise, recall was 3.6\%, 63.1\%, and 100\% for factors 1.10, 1.25, and 1.50; at 1\% noise it was 100\% for all factors.

Five of the 21 anchors accounted for 94 of the 150 injected spectra with robust full-spectrum errors above 10\%. Across all 21 anchors, error counts correlated with local uncertainty (Spearman $\rho=0.65$), measured as the linearized standard deviation of $\log R_c$ at 5\% component noise. This diagnostic uses a Gaussian least-squares Jacobian normalized by latent impedance magnitude, accounting for the other circuit parameters. The descriptive association supports examining parameter identifiability alongside flagged-point sensitivity.

\subsection{Measured Residuals and End-to-End Review}
Public two-RC fits recovered DC resistance, the sum of all three terminal resistances including the subepithelial contribution, with maximum relative error 4.42\% across 120 evaluation spectra. M1 flagged 17/97 public measured spectra using the transferred historical cutoff. These checks demonstrate source-model adaptation and flag transfer; fault calibration is specific to each measurement source.

M1 marked 233/282 local measured spectra and M5 marked 185; their union marked 268. The 906 union-marked points included 358 endpoints. Archived robust-fit $E_m$ had median 2.14\% and 95th percentile 4.08\%; two spectra exceeded 10\%, at 63.66\% and 67.19\% (Fig.~\ref{fig:local}). M1 also marked noiseless spectra from all 21 anchors, identifying trend approximation and residual normalization as calibration targets.

After matched robust fitting, center the relative residuals $d_k$ as $w_k=d_k-\bar d$, ordered by frequency. Their lag-1 correlation, $\operatorname{Re}(w_{1:N-1}^H w_{2:N})/(\|w_{1:N-1}\|_2\|w_{2:N}\|_2)$, had median 0.715 (interquartile range 0.571--0.831) across 282 measured spectra, versus $-0.044$ ($-0.101$--0.019) across 126 simulated controls. This structure motivates evaluation with correlated noise and model discrepancy.

For exp1\_0310, index 11, flags were $\{0,59,61,62,64\}$. Retaining endpoints and deleting internal flags $\{59,61,62\}$ changed matched robust $R_c$ from 427.716 to 428.005~$\Omega$ (0.068\%); full-fit $E_m$ was 2.48\%. The agent retained the full estimate and existing model/parameter annotations, recorded the sensitivity result, and requested paired-blank and acquisition context.

All 215 internal-flag checks produced shifts below 10\%; 53 further spectra had endpoint flags exclusively. Replay preserved existing parameter-review requests for 272 spectra and both severe-mismatch indications. The internal-deletion rule added zero escalations. These results establish consistent execution and preservation of review evidence.

\section{Practical Interpretation and Scope}
The paired results support robust full-spectrum estimation and matched sensitivity checks. Robust M-estimation reduces large-residual sensitivity~\cite{Huber1964}; the endpoint results support inspection before removal. A small deletion shift establishes stability under that check; anchor-level uncertainty motivates complementary identifiability checks, while model adequacy and calibration determine physical interpretation.

The results describe the specified Gaussian-noise/amplitude-spike grid on seven previously examined records, with fixed full-spectrum pilot scales and starts. Broader evaluation can cover complete refitting pipelines, correlated noise, phase artifacts, and new biological records. Cell identities, biological independence, and interventions await source confirmation; matched blanks, area normalization, and gain calibration connect branch resistance to barrier physiology. Measured flags remain review annotations pending fault adjudication.

Evidence, model, rule, and action interfaces support human-in-the-loop extensions. An exploratory, unblinded Codex audit of Qwen3.6-plus explanations identified evidence misinterpretations and unsupported claims in evidence-only and rule-assisted outputs. These findings guide source-linked explanation checks, independent human assessment, and prospective review-time measurement.

\section{Conclusion}
Within the tested perturbation grid, controlled deletion comparisons support robust full-spectrum estimation with flag-guided inspection and matched sensitivity analysis. The framework implements these actions in measured-data replay, connecting the location-dependent numerical findings to a traceable review policy.

\section*{Acknowledgment}
MesoViva SAS, France, supported this internship research. AI tools are used in coding and manuscript preparation.

\clearpage
\IEEEtriggeratref{9}
\bibliographystyle{IEEEtran}
\bibliography{references}
\end{document}